\documentclass{aa}
\usepackage[utf8]{inputenc}

\usepackage[varg]{txfonts}
\usepackage{longtable,lscape}

\def\mstar{$M_{\rm *}$}

\def\arcsec{\hbox{$^{\prime\prime}$}}

\def\micron{$\mu$m}

\def\ergscm{ergs s$^{-1}$ cm$^{-2}$}

\def\msun{M$_{\odot}$}
\def\msunpyr{M$_{\odot}$ yr$^{-1}$}

\def\mstar{M$_*$}

\def\d4000{$D_{\rm 4000}$}
\def\halpha{\ifmmode {\rm H{\alpha}} \else $\rm H{\alpha}$\fi}
\def\hbeta{\ifmmode {\rm H{\beta}} \else $\rm H{\beta}$\fi}
\def\hgamma{\ifmmode {\rm H{\gamma}} \else $\rm H{\gamma}$\fi}
\def\hda{\ifmmode {\rm H{\delta}_{\rm A}} \else $\rm H{\delta}_{\rm A}$\fi}
\def\hdelta{\ifmmode {\rm H{\delta}} \else $\rm H{\delta}$\fi}
\def\oii{[O\,{\sc ii}]$\lambda$3727}

\def\oiiib{[O\,{\sc iii}]$\lambda$5007}

\def\niib{[N\,{\sc ii}]$\lambda$6584}
\def\nii{[N\,{\sc ii}]$\lambda\lambda$6548,6584 }
\def\sii{[S\,{\sc ii}]$\lambda\lambda$6717,6731}

\def\neiii{[Ne\,{\sc iii}]$\lambda$3869}

\def\niisha{[N\,{\sc ii}]/H$\alpha$}

\def\nn2{$N2$}
\def\rr23{$R_{\rm 23}$}
\def\oo32{$O_{\rm 32}$}
\def\oo2ne3{$O_{\rm 2Ne3}$}
\begin{document}
\title{MASSIV: Mass Assembly Survey with SINFONI in VVDS\thanks{This work is based  
    on observations collected at the European Southern Observatory
    (ESO) Very Large Telescope, Paranal, Chile, as part of the Programs 
    179.A-0823, 78.A-0177, and 75.A-0318. This work also benefits from data products 
    produced at TERAPIX and the Canadian Astronomy Data Centre as part of the 
    Canada-France-Hawaii Telescope Legacy Survey, a collaborative project of NRC 
    and the CNRS. 
}}
    \subtitle{VI. Metallicity-related fundamental relations in star-forming galaxies at $1 < z < 2$}

\author{C. Divoy\inst{\ref{inst1}}$^{,}$\inst{\ref{inst1-2}}
\and T. Contini \inst{\ref{inst1}}$^{,}$\inst{\ref{inst1-2}}
\and E. P\'erez-Montero\inst{\ref{inst3}}
\and J. Queyrel \inst{\ref{inst1}}$^{,}$\inst{\ref{inst1-2}}
\and B. Epinat\inst{\ref{inst4}}
\and C. L\' opez-Sanjuan\inst{\ref{inst5}}
\and D. Vergani \inst{\ref{inst6}}
\and J. Moultaka \inst{\ref{inst1}}$^{,}$\inst{\ref{inst1-2}}
\and P. Amram\inst{\ref{inst4}}
\and B. Garilli\inst{\ref{inst7}}
\and M. Kissler-Patig \inst{\ref{inst8}}
\and O. Le Fèvre\inst{\ref{inst4}}
\and L.Paioro \inst{\ref{inst7}}
\and L.A.M. Tasca\inst{\ref{inst4}}
\and L. Tresse\inst{\ref{inst4}}
\and V. Perret\inst{\ref{inst4}}
} 

\offprints{C. Divoy, \email{Claire.Divoy@irap.omp.eu}}

\institute{
Institut de Recherche en Astrophysique et Plan\'etologie (IRAP), CNRS, 14 avenue Edouard Belin, F-31400, Toulouse, France \label{inst1}
\and IRAP, Universit\'e de Toulouse, UPS-OMP, Toulouse, France \label{inst1-2}
\and Instituto de Astrof\'\i sica de Andaluc\'\i a - CSIC Apdo. 3004 E-18080, Granada, Spain \label{inst3}
\and Aix Marseille Universit\'e, CNRS, LAM (Laboratoire d'Astrophysique  de Marseille) UMR 7326, 13388, Marseille, France  \label{inst4} 
\and Centro de Estudios de F\' isica del Cosmos de Arag\'on, Plaza San Juan 1, planta 2, E-44001, Teruel, Spain \label{inst5}
\and INAF-IASFBO, Via P. Gobetti 101, I-40129 Bologna, Italy \label{inst6}
\and IASF-INAF, via Bassini 15, I-20133, Milano, Italy \label{inst7}
\and Gemini Observatory, 670 N. A'ohoku Place, Hilo, Hawaii, 96720, USA \label{inst8}
} 

\date{Received 31 March 2014  / 
Accepted 19 June 2014 }

\abstract {} {
The MASSIV (Mass Assembly Survey with SINFONI in VVDS) project aims at finding constraints on the different processes involved in galaxy evolution. This study proposes to improve the understanding of the galaxy mass assembly through chemical evolution using the metallicity as a tracer of the star formation and interaction history.} {We analyse the full sample of MASSIV galaxies for which a metallicity estimate has been possible, that is 48 star-forming galaxies at $z\sim 0.9-1.8$, and compute the integrated values of some fundamental parameters, such as the stellar mass, the metallicity and the star formation rate (SFR). The sample of star-forming galaxies at similar redshift from zCOSMOS \citep{2012PerezMontero} is also combined with the MASSIV sample. We study the cosmic evolution of the mass-metallicty relation (MZR) together with the effect of close environment and galaxy kinematics on this relation.
We then focus on the so-called fundamental metallicity relation (FMR) proposed by \cite{2010Mannucci} and other relations between stellar mass, SFR and metallicity as studied by \cite{2010LaraLopez}. We investigate if these relations are really fundamental, i.e.  if they do not evolve with redshift. 
} {The MASSIV galaxies follow the expected mass-metallicity relation for their median redshift. We find however a significant difference between isolated and interacting galaxies as found for local galaxies: interacting galaxies tend to have a lower metallicity. The study of the relation between stellar mass, SFR and metallicity gives such large scattering for our sample, even combined with zCOSMOS, that it is difficult to confirm or deny the existence of a fundamental relation. 
} {}
\authorrunning{C. Divoy et al.}
\titlerunning{MASSIV: Integrated metallicities and fundamental relations}

\maketitle

\section{Introduction}
The current paradigm for galaxy formation is the so-called hierarchical model. It explains the formation of small galaxies at the earliest epochs, mainly by gas accretion in dark matter haloes. Later on in the history of the Universe,  these small galaxies merge to produce larger galaxies. During these processes the interstellar medium (ISM) undergoes multiple mixings, inducing strong variations in the chemical abundances of galaxies. The abundance of metals in the ISM defined as the metallicity ($Z$), is thus a marker of the galaxy star formation history and of the various gas flows (accretion and outflows) that could happen during its life. The metal enrichment of the ISM is due to stellar evolution but is also a consequence of galaxy interactions and/or interplay with the surrounding intergalactic medium. Indeed, both inflows of pristine gas toward the center of the galaxy, and outflows of metal-rich gas due to supernovae-driven galactic winds, reduce the nuclear metallicity of galaxies. Thereby metallicity is a fundamental parameter to follow along cosmic time for the understanding of the formation and evolution of galaxies.\

Simulations based on hierarchical model which include chemical evolution and feedback processes \citep{2012Moster,2012Calura,2013Stinson} can roughly reproduce observations but new and more accurate datasets are needed to constrain the free parameters of these simulations such as the feedback efficiency and accretion rate. For instance, \cite{2008MichelDansac} performed simulations to constrain the role of galaxy interactions in the variation of the metallicity. They found that interacting galaxies with low(high) masses have higher(lower) metallicities than their isolated counterparts following the well-known mass-metallicity relation. During minor mergers the metallicity of the lighter component increases between the start of the interaction and the coalescence but for a major merger the metallicity of the heavier component increases.

Since the seventies, the existence of a link between galaxy luminosity and metallicity is known. \cite{1979Lequeux} give a relation between these parameters for galaxies in the local universe. Nevertheless the luminosity is too sensitive to the star formation rate (SFR) and extinction so subsequent studies introduced a more fundamental parameter: the stellar mass. Large surveys of local galaxies, such as 2dF and SDSS, have been used to study and confirm the relation between stellar mass and metallicity \citep{2004Lamareille,2004Tremonti}, also called MZR. For instance, \cite{2004Tremonti} established the relation between stellar mass and metallicity for a sample of more than $50\,000$ local galaxies from the SDSS. The origin of this relation is not well understood yet. We know theoretically that 
stellar evolution ends with the ejection of metals that enrich the ISM with time, and that stellar mass increases also with time mainly by merger processes. As predicted by theory, numerous studies based on distant galaxies \citep[e.g.][]{2006Erb, 2006Lamareille, 2009Lamareille, 2008Maiolino, 2009Mannucci, 2009PerezMontero, 2012PerezMontero} show that metallicity decreases with redshift up to $z\sim 4$. However the role of inflows and outflows is not well constrained. Each process is theoretically well understood but their scale factors and their dependence on time are not well defined. For instance, at high redshift, gas inflows are necessary to reproduce unusual metallicity gradients in galaxies \citep[e.g.][]{2010Cresci, 2012Queyrel} but we do not know yet if these gas inflows are mainly due to cold flows or induced by galaxy interactions. At this point, another parameter seems to play a role in galaxy evolution: the star formation rate (SFR). \cite{2010Mannucci} proposed a relation between mass, metallicity and SFR, called  the Fundamental Metallicity Relation (FMR). Independently \cite{2010LaraLopez} found a fundamental plane in the space defined by these three parameters. The evolution with redshift of these relations is not well known yet and quite controverted. On one hand there are studies  \citep[e.g.][]{2012Cresci} finding a non-evolving FMR and on the other hand other studies find a MZR evolving with redshift even when they remove the dependence on the SFR \citep[e.g.][]{2012PerezMontero}.

In this paper, we use the MASSIV sample \citep{2012Contini} to investigate the MZR and the FMR in the redshift range $0.9 < z < 1.8$. MASSIV is representative of the star-forming galaxy population at these redshifts, corresponding to a critical cosmic period for galaxy evolution when galaxies change rapidly.  
The aim of this study is to contribute to our understanding of the physical processes involved in the early stages of galaxy evolution. First results on the MZR, metallicity gradients, and fundamental relations between mass, velocity, and size,  all based on the so-called ``first epoch'' MASSIV sample, have been published in \cite{2009Queyrel}, \cite{2012Queyrel} and \cite{2012Vergani} respectively.

This paper is organized as follows. In Section \ref{sect:2}, we present the galaxy sample obtained from the MASSIV survey and how we extract \halpha\ and \niib\ fluxes to compute the SFR and the metallicity of galaxies. In Section \ref{sect:mz}, we investigate the MZR for our sample and its relation with various parameters such as the redshift, the galaxy kinematics and their close environment. In Section \ref{sect:FMR} we present an analysis of the relations between stellar mass, metallicity and SFR for the MASSIV galaxies. In Section \ref{sect:conclusion} we discuss the agreement of our results with previous studies.

Throughout the paper, we assume a $\Lambda$CDM cosmology with $\Omega_m=0.3$, $\Omega_\Lambda=0.7$ and $H_0 = 70~ \mathrm{km s}^{-1}\mathrm{ Mpc}^{-1} $.

\section{Dataset}
\label{sect:2}
\subsection{The MASSIV sample}

For this analysis we use the full MASSIV (Mass Assembly Survey with SINFONI in VVDS) sample of 83 star-forming galaxies which is fully described in \cite{2012Contini}. Hereafter, we call as the ``first epoch'' sample the 50 galaxies first observed and published in \cite{2012Epinat}, and the ``second epoch'' sample the last 33 galaxies. These galaxies have been selected in the VVDS (VIMOS VLT Deep Survey \citep{2013LeFevre}) sample to be representative of the population of star-forming galaxies at high redshifts ($z\sim 0.9-1.8$). The selection of galaxies up to redshift $z \sim 1.5$ was based on the intensity and equivalent width of the \oii\ emission line measured on VIMOS spectra \citep[e.g.][]{2007Franzetti, 2008Vergani, 2009Lamareille}. The selection of galaxies at higher redshift was based mainly on their rest-frame ultraviolet continuum and/or massive stars absorption lines. These star formation criteria ensure that the rest-frame optical emission lines \halpha\ and \niib, or in a few cases \oiiib, are bright enough to be observed with the Integral Field Unit SINFONI in the near-infrared $J$ (sources at $z < 1.1$) and $H$ (sources at $z > 1.2$) bands. As shown in \cite{2012Contini}, the final MASSIV sample provides a good representation of ``normal'' star-forming galaxies at $0.9 < z < 1.8$ in the stellar mass regime $M_* = 10^9-10^{11}$ \msun, with a median star formation rate SFR $\sim 30$ \msunpyr\ and a detection threshold of $\sim 5$ \msunpyr\ . 

SINFONI observations have been performed between April 2007 and January 2011. Most (85\%) of the targeted galaxies have been observed in a seeing-limited mode (with a spatial sampling of $0.125 \arcsec$/pixel). Eleven galaxies have been acquired with adaptive optics (AO) assisted with a laser guide star (seven with $0.05\arcsec$ and four with $0.125\arcsec$/pixel spatial sampling). 

The data reduction was performed with the ESO SINFONI pipeline (version 2.0.0), using the standard master calibration files provided by ESO. The absolute astrometry for the SINFONI data cubes was derived from nearby bright stars also used for point spread function measurements. Custom IDL and Python scripts have been used to flux calibrate, align, and combine all the individual exposures. For each galaxy a non sky-subtracted cube was also created, mainly to estimate the effective spectral resolution. For more details on data reduction, we refer to \cite{2012Epinat}. 

\subsection{Lines Measurement}

\subsubsection{Extraction of 1D spectra}
\label{sect:1dspectra}

Spectra were extracted from both flux calibrated and counts datacubes on apertures corresponding to a signal-to-noise ratio ($SNR$) on the \halpha\ line larger than 2 and to spaxels where the line is narrower than the spectral resolution. $SNR$ maps were obtained using a $2\times 2$ spaxels Gaussian smoothing as described in \cite{2012Epinat}. Note that isolated extra-spaxels with a $SNR > 2$  were removed based on a visual inspection. Spectra of all the spaxels belonging to the aperture were shifted in wavelength according to their corresponding \halpha\ line position, and then summed to obtain the final 1D integrated spectrum.This shift enables to have a spectrum with narrower lines, better fitted by a Gaussian function than in the case where the velocity gradient is not removed. This procedure also increases the $SNR$ in the final integrated spectrum and therefore the accuracy of line measurements.

\subsubsection{Procedure to measure emission lines}

\begin{figure}[t]
\centering
\includegraphics[width=0.45\textwidth]{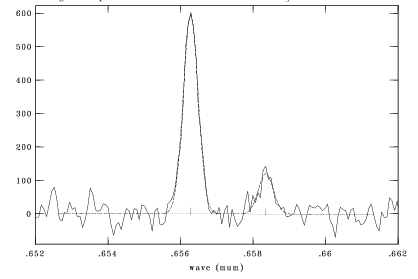}
\caption{One example of the measurement of \halpha\ and \niib\ emission lines with \textsc{splot} task of IRAF. The spectrum and the fitted gaussians are shown with solid and dotted lines respectively. Intensity units are expressed in number counts and rest-frame wavelength units in \micron .}
\label{fig:fit}
\end{figure}

To compute metallicities, we need to measure the flux of  \halpha\ and \niib\ emission lines. To do so we use the task \textsc{splot} of IRAF to fit Gaussian functions on the different lines of the 1D spectrum of each galaxy. We obtain the flux and associated errors with the deblending mode of this task \textsc{splot} with the constrain that the FWHM of the \halpha\ and \niib\  lines are the same (see Fig.\ref{fig:fit}).

For some galaxies the lines were not measurable for two main reasons: either the intensity of the lines is too low to be measured and so there is no detection for these lines, or there are some strong sky line residuals at the same wavelength as the expected lines. For the second epoch sample, there are seven galaxies with no detection of the  \niib\  emission line: VVDS020258016, 910154631, 910184233, 910187744, 910232719, 910296626, 910300117 and six galaxies with strong sky line residuals at the wavelength of \halpha\ or \niib\ emission lines: VVDS910159867, 910177382, 910191357, 910195040, 910274060, 910377628. Similar situation occurred for the first epoch sample and have been explained in \cite{2012Queyrel}.

\subsection{Metallicity}

We were able to derive the integrated metallicity for 48 galaxies of the sample. 
Metallicity is computed using the flux ratio between \halpha\ and \niib\ emission lines, the so-called N2 parameter. This parameter has a monotonic linear relation with oxygen abundance up to oversolar values and it has the advantage that it has almost no dependence with reddening or flux calibration. Although N2 is also related to ionization parameter and nitrogen-to-oxygen abundance ratio, the empirical calibration proposed by \cite{2009PerezContini}  based on objects with a direct determination of the electron temperature, leads to metallicities with an uncertainty around $0.3$ dex. For the identified interacting galaxy pairs \citep[see][]{2012Lopez}, the spectrum of the main component has been extracted and used to compute the metallicity. Depending on the calibration used, the derived metallicities can change significantly. We use the relation given by \cite{2008Kewley} and \cite{2009Queyrel} to convert metallicities computed with the \cite{2009PerezContini} calibration to metallicities using the \cite{2004Tremonti} calibration. This relation is:
\begin{equation}
y = - 0.217468 x^3 + 5.80493 x^2 - 50.2318 x + 149.836
\end{equation}
\label{sect:metallicite}
where $x$ or $Z_{PMC09}$ is the metallicity in \cite{2009PerezContini} calibration and $y$ or $Z_{T04}$ metallicity in \cite{2004Tremonti} calibration. Table~\ref{Table1} contains for each galaxy:  the galaxy VVDS identification number, the redshift as derived from the SINFONI data using the position of the \halpha\ line, the flux of the \halpha\  line, the flux ratio between \niib\ and \halpha\ lines, the metallicity computed with the \cite{2009PerezContini} calibration and the recalibration of this metallicity to \cite{2004Tremonti} calibration.

\begin{figure}[t]
\centering
\includegraphics[width=0.37\textwidth]{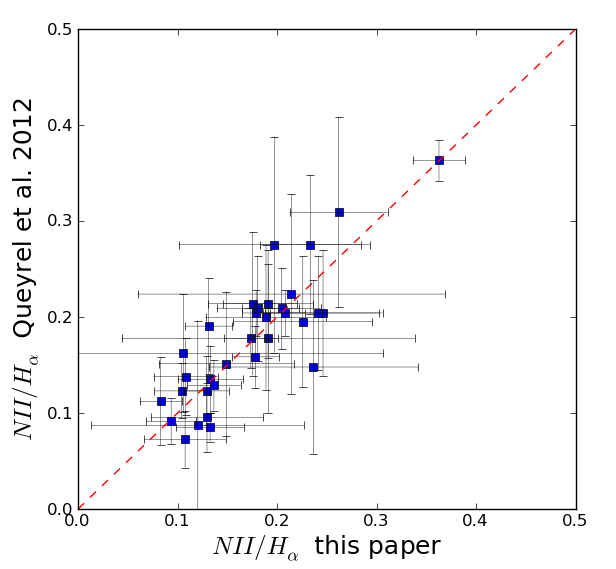}
\caption{Comparison between the \niisha\ flux ratios measured in this paper and those published in \cite{2012Queyrel}. The agreement is good when error bars, coming essentially from measurement uncertainties, are taken into account ($\chi ^2 = 0.648$).}
\label{fig:comp}
\end{figure}

The \halpha\ flux, the flux ratio \niisha, and the metallicity for 34 galaxies of the first epoch sample of MASSIV have already been published in \cite{2012Queyrel}. For the sake of homogeneity, we derive again these quantities for the full MASSIV sample using the same procedure as \cite{2012Queyrel}. Figure~\ref{fig:comp} shows the comparison of previous and new measurements for the \niisha\ flux ratio. Measurements are consistent with each other taking the uncertainties into account. Note that the error bars are mainly due to the measurement of the faint \niib\ line.

\subsection{Star formation rate and stellar mass}
\label{sect:SFR}

We derive the SFR from the \halpha\ flux measured in the integrated spectrum of the galaxies following the \cite{1998Kennicutt} relation. This SFR is also corrected for interstellar gas extinction. For each galaxy, the value of this extinction is an output of the SED fitting performed on extensive multi-band photometry \citep{2012Contini}. The method used to compute the SFR, the extinction-corrected SFR and their associated errors is explained in \cite{2009Epinat}. The median value of the errors on the SFR is $2.2$ \msunpyr .\

The stellar masses, which are also an output of the SED fitting procedure are those listed in \cite{2012Contini}. Values of SFR (before and after extinction correction) and stellar masses are listed in Table~\ref{Table1}.

Stellar masses and SFR have been computed using a \cite{1955Salpeter} initial mass function (IMF). For the sake of consistency and to allow comparison with previous studies, stellar masses and SFR for other datasets computed with a different IMF have been corrected according to the offsets given in \cite{2010Bernardi}. For instance, the relations from \cite{2004Tremonti}, \cite{2012Yabe} and \cite{2010LaraLopez} using \cite{2002Kroupa} IMF have been shifted in figures \ref{fig:relMZ}, \ref{fig:relMZ_mass}, \ref{fig:relMZ_rot}, \ref{fig:relMZ_int}, \ref{fig:FMRll} and \ref{fig:FMRll_us}. The relation of \cite{2010Mannucci} and stellar masses from \cite{2012PerezMontero} using \cite{2003Chabrier} IMF have been corrected in figures \ref{fig:FMR}, \ref{fig:FMR05_int} and \ref{fig:relMZ_PM}.

\begin{longtab}
\begin{landscape}
\begin{longtable}{ccrrrrrrr}
\caption{Fundamental parameters of the MASSIV sample. Column~1 lists the VVDS identification number. Column~2 gives the redshift. Column~3 and 4 list the \halpha\ flux (in \ergscm) and the flux ratio \niisha\ respectively. The star formation rate (in \msunpyr) is given in columns~5 (not corrected for extinction) and 6 (dereddened value). Columns~7 and 8 list the metallicity computed with the \cite{2009PerezContini} and \cite{2004Tremonti} calibrations respectively.  Stellar mass (in log of \msun) is given in column~9 (from \cite{2012Contini}).} \\
\hline\hline
VVDS ID & Redshift &  Flux(\halpha) & \niisha\ & SFR & SFR$_{\rm c}$ & $Z_{\rm PMC09}$  & $Z_{\rm T04}$ & log(\mstar) \\
(1) & (2) & (3) & (4) & (5) & (6) & (7) & (8) & (9) \\
		\hline
020106882& 1.3980 & $ 14.4\pm 2.4$& $0.23\pm 0.07$ & $ 13.3\pm 0.8$& $ 38.10\pm 2.24$ &$ 8.56\pm 0.11$ & $ 8.80\pm 0.11$& $ 9.99\pm 0.22$\\
020116027& 1.5259 & $ 15.6\pm 1.5$& $0.13\pm 0.06$ & $ 14.7\pm 3.6$& $ 42.20\pm 1.62$ &$ 8.37\pm 0.15$ & $ 8.55\pm 0.15$& $10.09\pm 0.23$\\
020147106& 1.5174 & $ 39.5\pm 4.0$& $0.08\pm 0.03$ & $ 54.3\pm 6.8$& $ 92.10\pm 1.62$ &$ 8.22\pm 0.11$ & $ 8.38\pm 0.11$& $10.10\pm 0.38$\\
020164388& 1.3532 & $ 24.4\pm 2.5$& $0.11\pm 0.03$ & $ 21.1\pm 1.1$& $ 46.50\pm 1.86$ &$ 8.31\pm 0.10$ & $ 8.48\pm 0.10$& $10.13\pm 0.31$\\
020193070& 1.0273 & $  6.8\pm 1.5$& $0.20\pm 0.10$ & $  3.0\pm 0.4$& $ 14.40\pm 1.58$ &$ 8.51\pm 0.17$ & $ 8.73\pm 0.17$& $10.15\pm 0.20$\\
020214655& 1.0369 & $ 24.2\pm 7.4$& $0.20\pm 0.04$ & $ 10.6\pm 0.5$& $ 51.70\pm 2.51$ &$ 8.53\pm 0.07$ & $ 8.76\pm 0.07$& $10.02\pm 0.16$\\
020218856& 1.3096 & $  8.1\pm 0.9$& $0.10\pm 0.04$ & $  6.3\pm 0.6$& $ 29.65\pm 0.57$ &$ 8.27\pm 0.14$ & $ 8.43\pm 0.14$& $ 9.70\pm 0.26$\\
020255799& 1.0352 & $  5.4\pm 1.5$& $0.21\pm 0.15$ & $  2.6\pm 0.3$& $ 12.80\pm 2.34$ &$ 8.54\pm 0.25$ & $ 8.77\pm 0.25$& $ 9.87\pm 0.16$\\
020261328& 1.5291 & $  9.0\pm 2.7$& $0.12\pm 0.11$ & $ 11.2\pm 1.0$& $ 19.00\pm 3.71$ &$ 8.34\pm 0.30$ & $ 8.52\pm 0.30$& $10.01\pm 0.20$\\
020283083& 1.2809 & $ 13.5\pm 2.5$& $0.19\pm 0.06$ & $ 10.3\pm 0.8$& $ 17.40\pm 2.34$ &$ 8.50\pm 0.11$ & $ 8.72\pm 0.11$& $10.05\pm 0.21$\\
020283830& 1.3949 & $ 11.1\pm 1.8$& $0.15\pm 0.07$ & $ 10.3\pm 1.0$& $ 50.40\pm 1.62$ &$ 8.42\pm 0.16$ & $ 8.62\pm 0.16$& $10.37\pm 0.17$\\
020363717& 1.3335 & $ 38.5\pm 4.2$& $0.13\pm 0.03$ & $ 31.5\pm 3.6$& $ 53.40\pm 2.04$ &$ 8.38\pm 0.08$ & $ 8.57\pm 0.08$& $ 9.68\pm 0.20$\\
020370467& 1.3330 & $ 15.9\pm 2.5$& $0.25\pm 0.06$ & $ 14.6\pm 2.5$& $ 71.20\pm 1.38$ &$ 8.59\pm 0.08$ & $ 8.84\pm 0.08$& $10.57\pm 0.14$\\
020386743& 1.0465 & $ 18.6\pm 2.2$& $0.10\pm 0.03$ & $  8.1\pm 0.8$& $ 39.40\pm 3.24$ &$ 8.29\pm 0.09$ & $ 8.46\pm 0.09$& $ 9.88\pm 0.21$\\
020461235& 1.0351 & $  8.7\pm 0.9$& $0.18\pm 0.04$ & $  3.3\pm 0.9$& $  9.60\pm 1.99$ &$ 8.47\pm 0.09$ & $ 8.68\pm 0.09$& $10.36\pm 0.14$\\
140083410& 0.9426 & $ 45.5\pm 4.3$& $0.13\pm 0.02$ & $ 16.8\pm 1.2$& $ 37.10\pm 4.17$ &$ 8.37\pm 0.09$ & $ 8.55\pm 0.09$& $10.07\pm 0.18$\\
140123568& 1.0016 & $  6.6\pm 0.8$& $0.24\pm 0.06$ & $  2.4\pm 0.5$& $ 11.80\pm 2.34$ &$ 8.58\pm 0.09$ & $ 8.83\pm 0.09$& $ 9.73\pm 0.40$\\
140137235& 1.0444 & $  9.2\pm 1.6$& $0.11\pm 0.20$ & $  4.4\pm 1.1$& $ 21.60\pm 3.71$ &$ 8.30\pm 0.25$ & $ 8.47\pm 0.25$& $10.07\pm 0.29$\\
140217425& 0.9758 & $118.1\pm 4.4$& $0.36\pm 0.03$ & $ 41.0\pm 1.9$& $200.00\pm 1.51$ &$ 8.72\pm 0.02$ & $ 9.01\pm 0.02$& $10.84\pm 0.17$\\
140258511& 1.2421 & $ 49.2\pm 2.2$& $0.26\pm 0.05$ & $ 30.0\pm 1.9$& $146.40\pm 3.02$ &$ 8.61\pm 0.06$ & $ 8.87\pm 0.06$& $10.80\pm 0.48$\\
140262766& 1.2813 & $ 10.6\pm 2.6$& $0.19\pm 0.15$ & $ 10.0\pm 3.8$& $ 10.00\pm 4.79$ &$ 8.50\pm 0.26$ & $ 8.72\pm 0.26$& $ 9.84\pm 0.43$\\
140545062& 1.0401 & $ 27.5\pm 3.4$& $0.17\pm 0.03$ & $ 13.4\pm 1.5$& $ 22.70\pm 1.44$ &$ 8.47\pm 0.05$ & $ 8.68\pm 0.05$& $10.60\pm 0.18$\\
220014252& 1.3097 & $ 51.9\pm 6.9$& $0.19\pm 0.05$ & $ 40.5\pm 2.5$& $197.50\pm 2.63$ &$ 8.50\pm 0.08$ & $ 8.72\pm 0.08$& $10.78\pm 0.21$\\
220015726& 1.3091 & $ 54.4\pm 1.8$& $0.21\pm 0.02$ & $ 37.0\pm 5.2$& $106.40\pm 1.44$ &$ 8.53\pm 0.03$ & $ 8.76\pm 0.03$& $10.77\pm 0.27$\\
220376206& 1.2440 & $ 73.5\pm 8.0$& $0.09\pm 0.03$ & $ 51.2\pm 3.9$& $249.70\pm 2.45$ &$ 8.26\pm 0.09$ & $ 8.42\pm 0.09$& $10.67\pm 0.27$\\
220386469& 1.0240 & $ 16.5\pm 6.3$& $0.11\pm 0.04$ & $  7.5\pm 1.1$& $ 36.50\pm 1.41$ &$ 8.31\pm 0.13$ & $ 8.48\pm 0.13$& $10.80\pm 0.16$\\
220397579& 1.0367 & $ 71.9\pm 9.4$& $0.13\pm 0.03$ & $ 29.4\pm 4.0$& $143.20\pm 3.24$ &$ 8.38\pm 0.09$ & $ 8.57\pm 0.09$& $10.23\pm 0.17$\\
220544103& 1.3970 & $ 50.8\pm 7.4$& $0.23\pm 0.05$ & $ 52.2\pm 1.4$& $117.50\pm 2.57$ &$ 8.57\pm 0.07$ & $ 8.81\pm 0.07$& $10.71\pm 0.27$\\
220544394& 1.0077 & $ 29.8\pm 3.8$& $0.14\pm 0.03$ & $ 10.3\pm 0.7$& $ 50.10\pm 2.40$ &$ 8.39\pm 0.07$ & $ 8.58\pm 0.07$& $10.34\pm 0.23$\\
220576226& 1.0217 & $ 30.2\pm 3.1$& $0.18\pm 0.01$ & $ 13.6\pm 0.7$& $ 66.30\pm 2.75$ &$ 8.48\pm 0.03$ & $ 8.69\pm 0.03$& $10.31\pm 0.23$\\
220578040& 1.0461 & $ 20.2\pm 1.6$& $0.18\pm 0.02$ & $  9.4\pm 0.9$& $ 20.80\pm 1.62$ &$ 8.48\pm 0.05$ & $ 8.69\pm 0.05$& $10.72\pm 0.16$\\
220584167& 1.4637 & $ 64.6\pm 2.9$& $0.13\pm 0.02$ & $ 68.6\pm 3.4$& $202.60\pm 1.78$ &$ 8.37\pm 0.06$ & $ 8.55\pm 0.06$& $11.21\pm 0.25$\\
910163602& 1.3263 & $ 15.9\pm 3.1$& $0.37\pm 0.05$ & $ 11.9\pm 2.0$& $ 54.35\pm 2.04$ &$ 8.73\pm 0.05$ & $ 9.03\pm 0.05$& $10.83\pm 0.12$\\
910186191& 1.5399 & $ 35.4\pm 6.0$& $0.12\pm 0.02$ & $ 39.1\pm 6.1$& $ 31.43\pm 6.10$ &$ 8.35\pm 0.06$ & $ 8.53\pm 0.06$& $11.27\pm 0.45$\\
910193711& 1.5523 & $ 33.0\pm 6.9$& $0.18\pm 0.04$ & $ 40.6\pm22.7$& $197.70\pm 2.04$ &$ 8.48\pm 0.08$ & $ 8.69\pm 0.08$& $ 9.99\pm 0.18$\\
910207502& 1.4656 & $  9.3\pm 1.2$& $0.26\pm 0.04$ & $ 10.3\pm 1.0$& $ 50.22\pm 1.03$ &$ 8.61\pm 0.05 $& $ 8.87\pm 0.05$& $10.31\pm 0.33$\\
910224801& 1.3201 & $ 10.7\pm 3.1$& $0.37\pm 0.05$ & $  8.8\pm 1.8$& $ 18.12\pm 1.77$ &$ 8.73\pm 0.04 $& $ 9.03\pm 0.04$& $ 9.96\pm 0.20$\\
910238285& 1.5405 & $  7.7\pm 3.0$& $0.16\pm 0.03$ & $ 14.6\pm 2.6$& $ 41.89\pm 2.63$ &$ 8.44\pm 0.07 $& $ 8.64\pm 0.07$& $10.92\pm 0.13$\\
910250031& 1.6609 & $ 28.4\pm 1.4$& $0.03\pm 0.02$ & $ 48.6\pm 5.6$& $139.84\pm 5.58$ &$ 7.92\pm 0.19 $& $ 8.09\pm 0.19$& $10.15\pm 0.20$\\
910259245& 1.5174 & $  4.9\pm 1.8$& $0.27\pm 0.15$ & $  5.6\pm 1.0$& $ 27.08\pm 0.98$ &$ 8.62\pm 0.20 $& $ 8.88\pm 0.20$& $10.79\pm 0.12$\\
910261247& 1.4262 & $ 18.8\pm 3.3$& $0.21\pm 0.02$ & $ 19.8\pm 1.1$& $ 96.78\pm 1.14$ &$ 8.54\pm 0.03 $& $ 8.77\pm 0.03$& $10.57\pm 0.14$\\
910266034& 1.5642 & $ 15.8\pm 4.0$& $0.08\pm 0.03$ & $ 20.4\pm 3.0$& $ 76.25\pm 2.95$ &$ 8.19\pm 0.14 $& $ 8.34\pm 0.14$& $10.07\pm 0.20$\\
910276733& 1.3390 & $  3.8\pm 0.9$& $0.16\pm 0.06$ & $ 11.0\pm 2.1$& $ 53.67\pm 2.11$ &$ 8.44\pm 0.12 $& $ 8.64\pm 0.12$& $10.14\pm 0.19$\\
910279515& 1.3983 & $ 14.2\pm 1.5$& $0.24\pm 0.11$ & $ 14.2\pm 2.0$& $ 69.00\pm 4.68$ &$ 8.58\pm 0.15$ & $ 8.83\pm 0.15$& $10.79\pm 0.14$\\
910279755& 1.3127 & $ 23.5\pm 1.4$& $0.15\pm 0.05$ & $ 16.3\pm 1.3$& $ 27.71\pm 1.29$ &$ 8.42\pm 0.10 $& $ 8.62\pm 0.10$& $ 9.96\pm 0.24$\\
910337228& 1.3972 & $ 15.8\pm 1.2$& $0.09\pm 0.08$ & $ 39.1\pm16.6$&$ 44.97\pm 16.57$ &$ 8.23\pm 0.30 $& $ 8.39\pm 0.30$& $ 9.96\pm 0.20$\\
910340496& 1.3981 & $  4.5\pm 0.6$& $0.26\pm 0.13$ & $  3.7\pm 0.5$& $ 17.95\pm 0.50$ &$ 8.60\pm 0.18 $& $ 8.85\pm 0.18$& $10.03\pm 0.25$\\
910370574& 1.6632 & $ 12.6\pm 1.5$& $0.10\pm 0.02$ & $ 16.8\pm 3.6$& $ 82.00\pm 3.65$ &$ 8.28\pm 0.08 $& $ 8.44\pm 0.08$& $10.09\pm 0.24$\\
	\hline
VVDS ID & Redshift &  Flux(\halpha) & \niisha\ & SFR & SFR$_{\rm c}$ & $Z_{\rm PMC09}$  & $Z_{\rm T04}$ & log(\mstar) \\
		\hline\hline
		\\

\label{Table1}

\end{longtable}
\end{landscape}

\end{longtab}
%

\section{The mass-metallicity relation}
\label{sect:mz}

\cite{2004Tremonti} give an expression of the relation between stellar mass and metallicity for a large sample of star-forming galaxies in the local universe:
\begin{equation}
Z = -0.08026 m^2 + 1.847 m -1.492
\end{equation}
where $m= \log(M_*/M_\odot)$ and $Z=12+\log(O/H)$ .\\
The understanding of the shape of this relation is crucial to constrain the role of the various physical processes, such as outflows and gas accretion, involved in galaxy evolution. It is also interesting to understand the causes for the metallicity dispersion around the median relation. Indeed, for a given stellar mass, a galaxy can have different gas-to-star mass ratio, SFR, or merger history.  These parameters influence the value of the metallicity and can explain the dispersion of the galaxies around the relation found by  \cite{2004Tremonti}. At high redshifts ($z>3$), star-forming galaxies seem to follow a similar relation but at lower metallicity \citep{2008Maiolino}.

\begin{figure}[b!]
\centering
\includegraphics[width=0.45\textwidth]{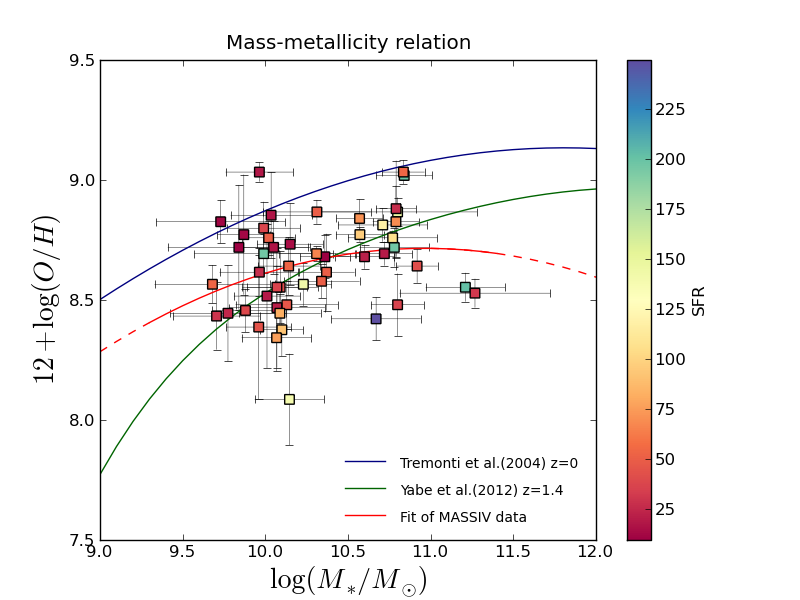}
\caption{Mass-metallicity relation. MASSIV galaxies are represented by coloured squares. The colour of the square scales with the value of the SFR. The red line is the best fit to the MASSIV sample. The mass-metallicity relations in the local universe (blue line) and at $z\sim 1.4$ (green line) are shown for comparison. Metallicities are given in the \cite{2004Tremonti} calibration.}
\label{fig:relMZ}
\end{figure}

With the MASSIV sample, we can put constraints on the mass-metallicity relation in the redshift range $z\sim 0.9-1.8$.  This relation can be compared with the one derived by \cite{2012Yabe} at redshift $z \sim 1.4$.  To compute the metallicities, \cite{2012Yabe} used the \cite{2004Pettini} calibration of the N2 parameter. To have a better comparison with our data (see Fig.~\ref{fig:relMZ}), we convert their metallicities into \cite{2004Tremonti} calibration with the relation given in \cite{2008Kewley}: 
\begin{equation}
y= -738.1193 + 258.9673 x -30.05705 x^2 + 1.167937 x^3
\end{equation}
where $x$ is the metallicity in \cite{2004Pettini} calibration and $y$ metallicity in \cite{2004Tremonti} calibration.

Figure~\ref{fig:relMZ} shows that MASSIV galaxies are distributed around the \cite{2012Yabe} mass-metallicity relation. The best fit of MASSIV galaxies, resulting in a scattering of $0.2$dex, with a second order polynomial relation (shown as a red line in  Fig.~\ref{fig:relMZ}) has the following expression:
\begin{equation}
Z= -0.11 m^2 + 2.45 m -4.71
\end{equation}

where $m= \log(M_*/M_\odot)$ and $Z = 12+\log(O/H)$. This fit has a $\chi ^2$ of $0.850$ and has been performed with the method of the least square taking into account the errors on both metallicity and mass.\
The agreement with either the \cite{2012Yabe} relation ($\chi ^2 = 0.678$) or the  \cite{2004Tremonti} relation offseted to lower metallicities ($\chi ^2 = 0.575$) is slighty worse.

\subsection{Median values}
\label{sect:median}

\begin{figure}[t!]
\centering
\includegraphics[width=0.45\textwidth]{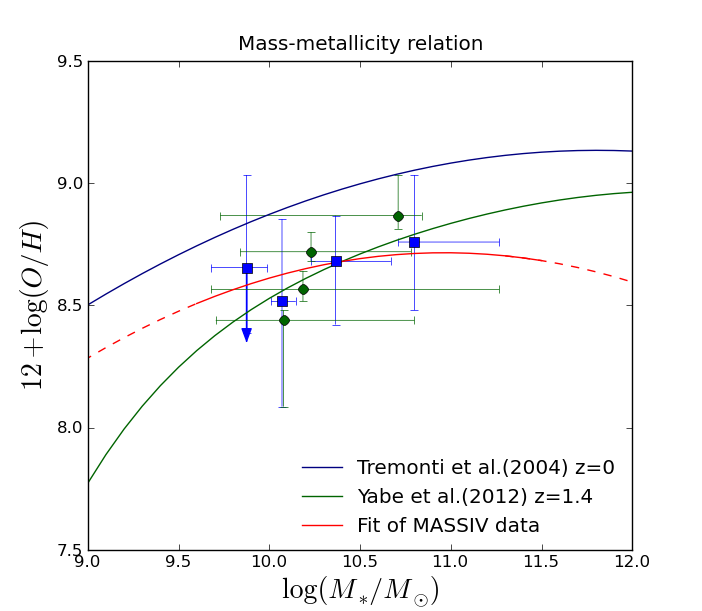}
\caption{Mass-metallicity relation with median values of stellar mass and metallicity for the MASSIV sample for a binning in mass (blue squares) and for a binning in metallicity (green dots). Error bars represent the dispersion around the median values. The arrow downward indicates that the median value of the metallicity for the lowest mass bin is certainly an upper limit due to incompleteness. Curves are the same as in Fig.~\ref{fig:relMZ}.}
\label{fig:relMZ_mass}
\end{figure}

The scatter around the mass-metallicity relation of MASSIV galaxies is quite important. It is thus interesting to compute the median value of metallicity and mass for different bins of stellar mass and bins of metallicity. We estimate the median value of metallicities and masses for 4 bins of mass and 4 bins of metallicity chosen to contain about the same number of galaxies in each bin. The median values are listed in Table~\ref{tab:median}. The blue squares in figure~\ref{fig:relMZ_mass} show the median values of the metallicity for a binning in mass and the green dots the median values of the stellar mass for a binning in metallicity. These two sets of median values show that the mass-metallicity relation for the MASSIV sample is in good agreement with  the \cite{2012Yabe} relation.\

\begin{table}[htdp]
\caption{Median values of stellar mass and metallicity for the MASSIV sample separated in bins of mass and in bins of metallicity.}
\begin{center}
\begin{tabular}{l r r}
\hline\hline
Stellar mass range & $M_{\rm median}$ & $Z_{\rm median}$ \\
\hline
$\log(M_* /M_{\odot}) \leq 10.00 $ & $9.87^{+0.11}_{-0.19}$ & $8.62^{+0.38}_{-0.27}$ \rule[-7pt]{0pt}{18pt}\\
$10.00 < \log(M_* /M_{\odot}) \leq 10.15$ & $10.07^{+0.08}_{-0.06}$ & $8.52^{+0.33}_{-0.43}$\rule[-7pt]{0pt}{18pt}\\
$10.15 < \log(M_* /M_{\odot}) \leq 10.70$ & $10.36^{+0.30}_{-0.13}$ & $8.68^{+0.91}_{-0.26}$\rule[-7pt]{0pt}{18pt}\\
$\log(M_* /M_{\odot}) \geq 10.70$ & $10.8^{+0.47}_{-0.09}$ & $8.76^{+0.27}_{-0.28}$\rule[-7pt]{0pt}{18pt}\\
\hline \hline
Metallicity range & $M_{\rm median}$ & $Z_{\rm median}$ \\
\hline
$Z \leq 8.49 $ & $10.08^{+0.72}_{-0.37}$ & $8.44^{+0.04}_{-0.35}$ \rule[-7pt]{0pt}{18pt}\\
$8.49< Z \leq 8.65$ & $10.19^{+1.08}_{-0.51}$ & $8.57^{+0.08}_{-0.05}$\rule[-7pt]{0pt}{18pt}\\
$8.65< Z \leq 8.81$ & $10.23^{+0.55}_{-0.39}$ & $8.72^{+0.08}_{-0.04}$\rule[-7pt]{0pt}{18pt}\\
$Z \geq 8.81$ & $10.71^{+0.13}_{-0.98}$ & $8.87^{+0.16}_{-0.06}$\rule[-7pt]{0pt}{18pt}\\
\hline
\end{tabular}
\end{center}
\label{tab:median}
\end{table}

The relatively high value of the median metallicity for the lowest mass bin is certainly due to incompleteness in this mass range. Indeed, the metallicity has been computed for $55\%$ of the MASSIV galaxies in this mass bin whereas metallicity has been derived for more than $80\%$ in higher mass bins. 
The median metallicity for $\log(M_* /M_{\odot}) \leq 10$ galaxies should be thus interpreted as an upper limit as low-metallicity galaxies in this low-mass range are difficult to probe due to the faintness of the \niib\ emission line.\

The median value of metallicity for the highest mass bin seems to be off the MZR  fitted on the full sample (red line) because, at high mass, this fit is dominated by two massive galaxies with a quite low metallicity. However the trend toward lower metallicities for the high mass end of the MZR is also present when combining the MASSIV galaxies with the zCOSMOS sample (see sect. \ref{sect:zcosmos} and Fig. \ref{fig:relMZ_PM} ).

\subsection{Is there a link with kinematics and/or galaxy interactions?}

\begin{figure}[ht!]
\centering
\includegraphics[width=0.45\textwidth]{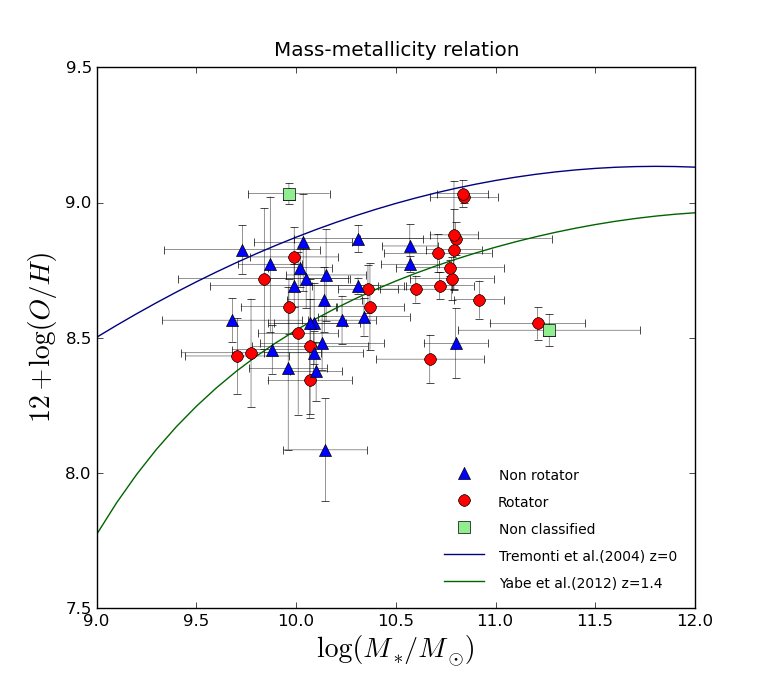}
\caption{Mass-metallicity relation for rotating (red dots), and non-rotating (blue triangles) MASSIV galaxies. The non-classified galaxies are shown as green squares. Curves are the same as in Fig.~\ref{fig:relMZ}.}
\label{fig:relMZ_rot}
\end{figure}

Thanks to SINFONI IFU observations, a first order kinematics classification of MASSIV galaxies has been performed by \cite{2012Epinat} and Epinat et al.\ (in prep.). The galaxies with a velocity map showing clear velocity gradients are classified as rotating disks whereas the other ones are classified as non-rotators. Figure~\ref{fig:relMZ_rot} shows that there is no significant difference in the mass-metallicity relation between the two samples. The rotators have a median metallicity of $8.68\pm0.19$ with a median mass of $10.09\pm0.11$ and non-rotators of $8.64\pm0.19$ with a median mass of $10.64\pm0.18$. The main difference between the two samples is that the median mass of the rotators is higher than the median mass of non-rotators and almost all the galaxies with masses greater than $10^{10.5}$\msun\ are rotators.\\

\begin{figure}[ht!]
\centering
\includegraphics[width=0.45\textwidth]{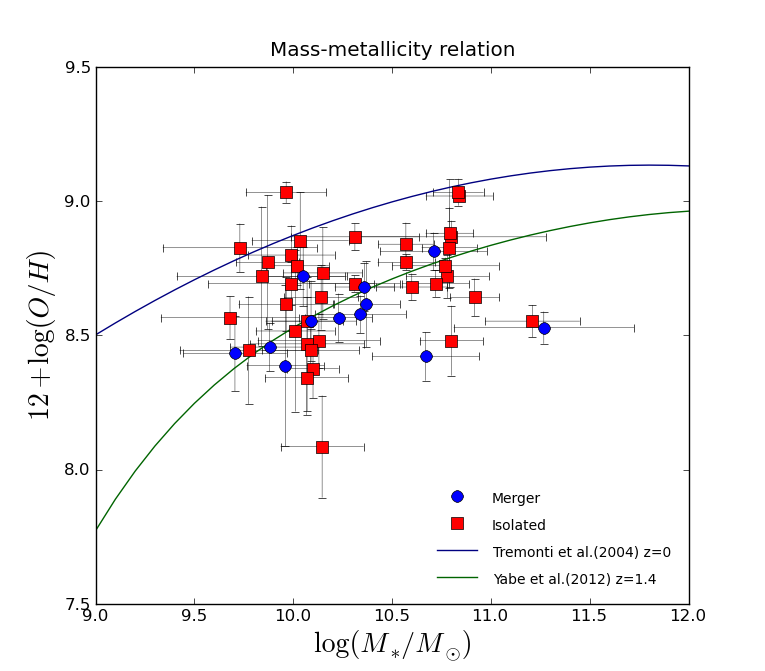}
\caption{Mass-metallicity relation for interacting (blue dots) and isolated (red squares) MASSIV galaxies. See Fig.~\ref{fig:relMZ} for lines description.}
\label{fig:relMZ_int}
\end{figure}

The metallicty of a galaxy could be affected by its close environment \citep[e.g.][]{2008MichelDansac}. An accurate classification of the level of gravitational interaction of MASSIV galaxies is given in  \cite{2012Lopez}. Galaxies are classified in two main groups: interacting galaxies (major or minor mergers) and isolated galaxies. Figure~\ref{fig:relMZ_int} shows clearly that interacting galaxies in the MASSIV sample have lower metallicities (with a median value of $8.56\pm0.12$ and a median mass of $10.28\pm0.17$) than isolated galaxies which have a median value of metallicity of $8.69\pm0.20$ and a median mass of $10.14\pm0.16$.
\section{Relations between stellar mass, metallicity and SFR}
\label{sect:FMR}

\cite{2010Mannucci} proposed a more fundamental relation between the stellar mass, SFR and metallicity of star-forming galaxies in the local universe, a relation called Fundamental Metallicity Relation (FMR). This relation does not seem to evolve with redshift at least up to $z \sim 2.5$. To visualize this relationship more easily, they propose a projection that remove the secondary dependences: 
\begin{equation}
\mu _{\alpha} = \log(M_*/M_{\odot}) - \alpha \log(SFR)
\label{eq:fmr}
\end{equation}

\begin{figure}[t!]
\centering
\includegraphics[width=0.4\textwidth]{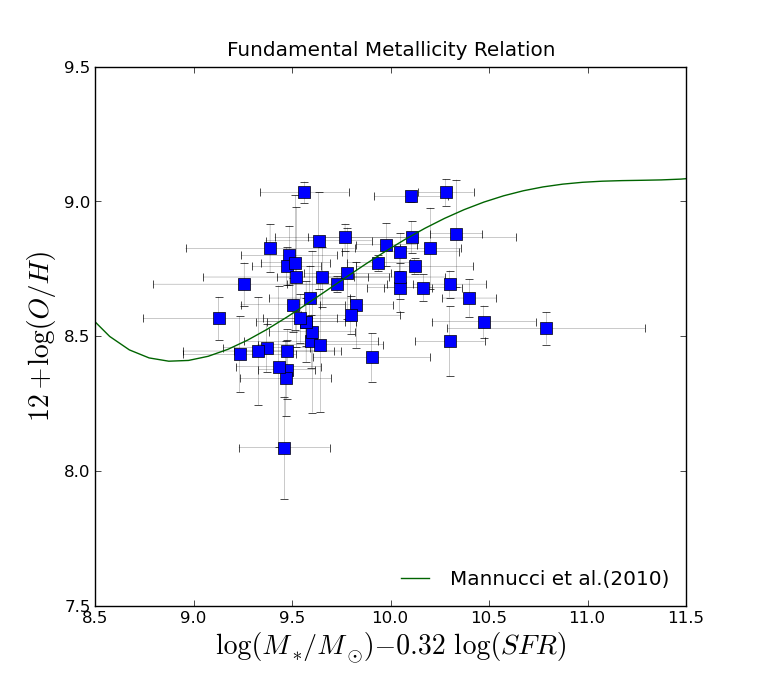}
\caption{Fundamental Metallicity Relation for MASSIV galaxies (blue squares) and assuming $\alpha = 0.32$ in Eq. \ref{eq:fmr}. The FMR defined by \cite{2010Mannucci} for the star-forming galaxies in the local universe is shown as the green curve.}
\label{fig:FMR}
\end{figure}

\cite{2010Mannucci} found a value of $\alpha = 0.32$ that minimizes the scattering of galaxies around the FMR in the local universe, with a small residual dispersion of $\sim 0.05$ dex in metallicity, i.e. $\sim 12$ per cent. Figure~\ref{fig:FMR} shows the FMR, i.e. the metallicity as a function of $\mu_{0.32}$, for the MASSIV sample. The metallicity has been (re)computed with the \cite{2004Tremonti} calibration as explained in section~\ref{sect:metallicite}, as well as the FMR from \cite{2010Mannucci} to account for different IMF (see section ~\ref{sect:SFR}). 
The intrinsic scatter of MASSIV galaxies around this relation is about $0.20$ dex.

%
%

\begin{figure}[hb!]
\centering
\includegraphics[width=0.4\textwidth]{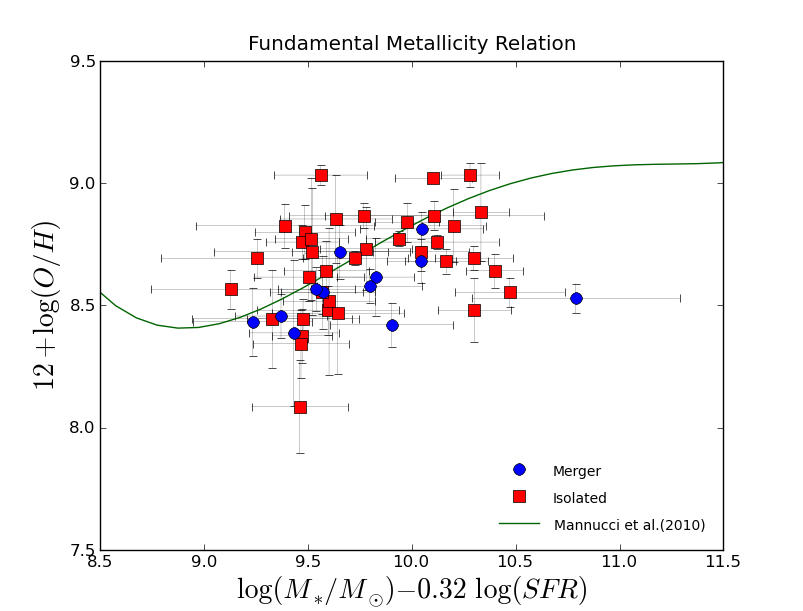}
\caption{Fundamental Metallicity Relation for MASSIV interacting (blue dots) and isolated (red squares) galaxies assuming $\alpha =0.32$ in Eq. \ref{eq:fmr}.}
\label{fig:FMR05_int}
\end{figure}

\begin{figure}[t!]
\centering
\includegraphics[width=0.35\textwidth]{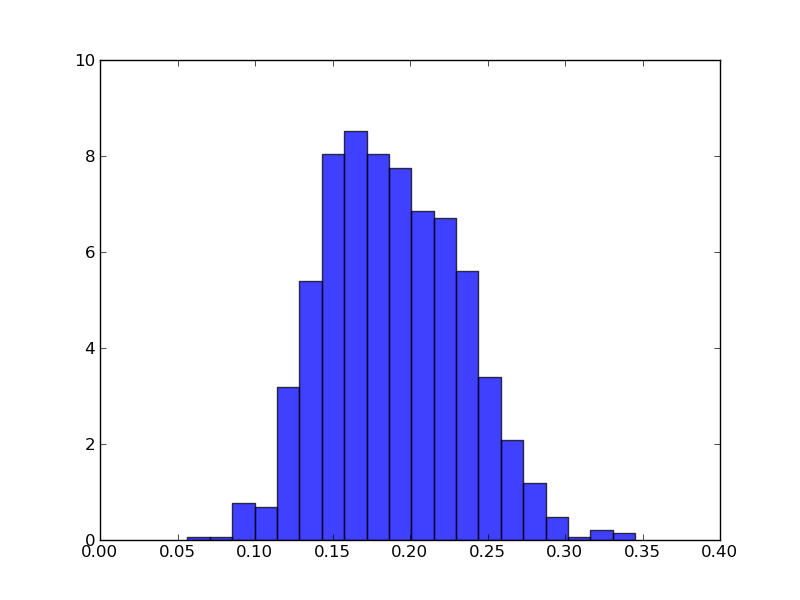}
\caption{Distribution of the scatter around the FMR for 12 isolated galaxies chosen randomly among the 37 isolated galaxies of the MASSIV sample.}
\label{fig:histo_disp}
\end{figure}

Figure~\ref{fig:FMR05_int} shows the FMR for MASSIV interacting and isolated galaxies. As for the mass-metallicity relation, there seems to be a significant difference between interacting and isolated galaxies. The intrinsic scattering around the FMR is about $0.12$ dex for interacting galaxies and $0.20$ dex for isolated ones. However, this difference could be due to the different size of the two sub-samples. To test this effect, we pick up randomly 12 isolated galaxies among the 37 isolated ones of the MASSIV sample. This trial is repeated 1000 times in order to compute the scattering of these smaller samples, with a size similar to the interacting sample. The resulting median value of the scatter of these 1000 smaller samples is $0.18$ dex, a value similar to the one measured in the observed sample of 37 isolated galaxies (see distribution in Fig.~\ref{fig:histo_disp}). We thus conclude that the difference of scattering between isolated and interacting MASSIV galaxies is significant. As for the mass-metallicity relation we do not find any difference between rotating disks and non-rotators. 

\cite{2010LaraLopez} also studied the fundamental plane for star-forming galaxies of the local universe using the SDSS sample. They propose a projection where the stellar mass is fixed. This approach is motivated by the fact that the stellar mass is the most fundamental parameter of galaxies, with an evolution time-scale longer than the SFR and metallicity. They thus defined a combination of the two other parameters (metallicity and SFR) as follows:
\begin{equation}
\log(M_* / M_\odot) = \alpha [12+\log(O/H)] + \beta \log(SFR) + \gamma
\label{eq:fmr_ll}
\end{equation}  

\cite{2010LaraLopez} find a parametrisation that minimizes the scattering around this relation at $0.16$ dex when $\alpha = 1.122\pm0.008$, $\beta = 0.474\pm0.004$ and $\gamma = -0.097\pm0.077$.  In their most recent paper, \cite{2013LaraLopez} find other values for these parameters: $\alpha = 1.3764\pm0.006$, $\beta = 0.6073\pm0.002$ and $\gamma = -2.5499\pm0.058$. This new study includes galaxies up to redshift 0.36 from the Galaxy and Mass Assembly (GAMA) survey in addition to the data from the SDSS-DR7 already studied in their previous paper  \citep{2010LaraLopez}.\

\begin{figure}[t!]
\centering
\includegraphics[width=0.4\textwidth]{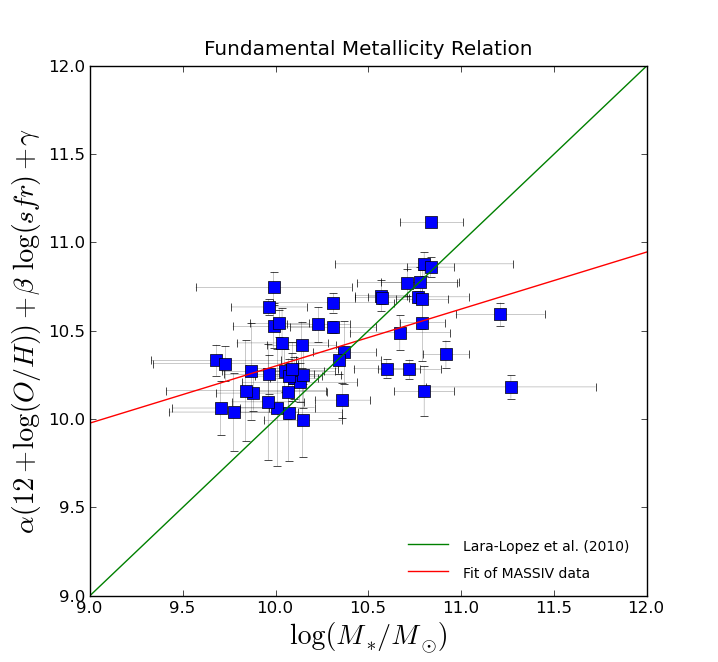}
\caption{Fundamental Metallicity Relation as defined by \cite{2010LaraLopez}, see Eq. \ref{eq:fmr_ll}. MASSIV galaxies are the blue squares.}
\label{fig:FMRll}
\end{figure}

Figure~\ref{fig:FMRll} shows this relation for the MASSIV sample. We find {\it an intrinsic} scatter of $0.26$ dex around the relation from \cite{2010LaraLopez}, which is slightly larger than the one we found with the FMR relation of \cite{2010Mannucci}. However our data do not follow the relation given by these values of $\alpha$, $\beta$ and $\gamma$ (green line in Fig. \ref{fig:FMRll}). MASSIV data are better fitted with the following relation (red line in Fig. \ref{fig:FMRll}): 
\begin{equation}
\alpha [12+\log(O/H)] + \beta \log(SFR) + \gamma = 0.323 \log(M_*/ M_\odot) + 7.071
\end{equation}
To derive these values, we performed a linear regression using least square method weighted on the errors of the three parameters.\

\begin{figure}[h!]
\centering
\includegraphics[width=0.4\textwidth]{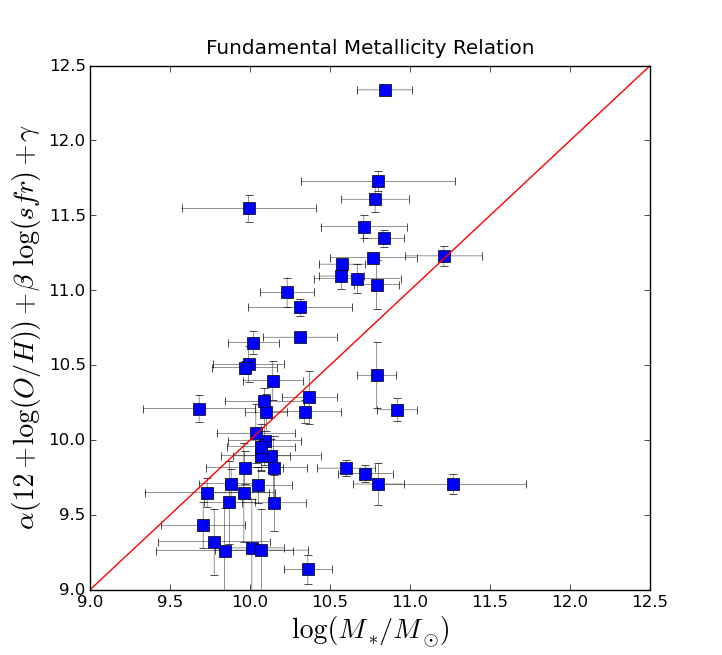}
\caption{Fundamental Metallicity Relation (red line) as defined by \cite{2010LaraLopez} and fitted to the MASSIV data in order to minimize the scatter}
\label{fig:FMRll_us}
\end{figure}

We thus tried to minimize the \textit{intrinsic} scattering of the MASSIV data around a relation with the same shape as proposed by \cite{2010LaraLopez} and found the following values for the parameters: $\alpha = 2.402$, $\beta = 1.811$ and $\gamma = -22.675$. These values of the parameters give a scatter of $0.15$ dex which is smaller than the scattering obtained with the other relations (see Fig.~\ref{fig:FMRll_us}). 

\newpage
\subsection{Comparison with zCOSMOS 20k sample}
\label{sect:zcosmos}

As shown previously, there is no convincing evidence that MASSIV galaxies do follow, i.e. with a reasonable small scatter, the so-called "fundamental" metallicity relations as defined by \cite{2010Mannucci} and \cite{2010LaraLopez}. In order to check if there is any bias in the MASSIV sample, we can compare our dataset with another sample at similar redshift. \cite{2012PerezMontero} performed an exhaustive study of the redshift evolution of the mass-metallicity relation for a large sample of star-forming galaxies in the local and intermediate-redshift Universe ($z = 0.02 - 1.32$) from the SDSS and zCOSMOS-20k datasets. This sample contains more than 40 galaxies at  $1< z<1.32$ that can be compared to the MASSIV sample. 

\begin{figure}[t!]
\centering
\includegraphics[width=0.45\textwidth]{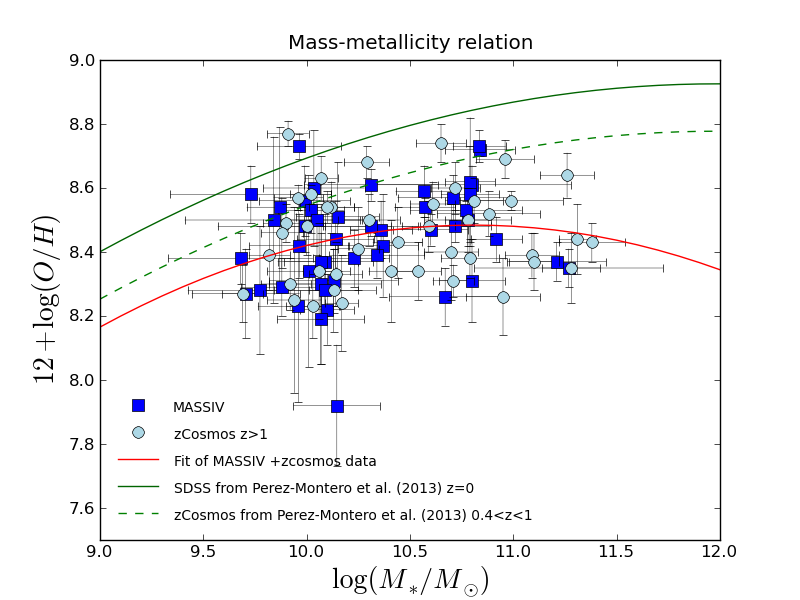}
\caption{Mass-metallicity relation. MASSIV galaxies are represented by blue squares. zCOSMOS galaxies are represented by light blue circles. The red line is the best fit to the MASSIV and zCosmos sample. The mass-metallicity relations in the local universe (green line) and at $0.4<z< 1$ (green dashed line) are shown for comparison. Metallicities are given in the \cite{2009PerezContini} calibration.}
\label{fig:relMZ_PM}
\end{figure}

Figure~\ref{fig:relMZ_PM} shows clearly that there is no significant difference between the location of the two samples in the mass-metallicity plane. Note that corrections have been applied to the zCOSMOS sample in order to comply with the metallicity calibration and IMF used for the MASSIV sample. We thus fit together the MASSIV data combined with the 43 galaxies from zCOSMOS. We found a fit similar to the one derived for the MASSIV sample alone. An evolution of the mass-metallicity relation with redshift is clearly seen in Fig.~\ref{fig:relMZ_PM}. Selection effects and incompleteness for the MASSIV sample are discussed in sect.~\ref{sect:median}. The zCOSMOS sample has the same minimal selection as the VVDS sample (parent sample for MASSIV), i.e. a lower limit in apparent magnitude combined with a flux-limited selection of star-forming galaxies. The main difference between MASSIV and zCOSMOS is however the method (or emission lines) used to compute metallicity. It is based on the \niisha\ ratio for MASSIV galaxies but makes use of \oii\ and \neiii\ emission lines for the zCOSMOS galaxies at $1< z<1.32$. Figure~\ref{fig:relMZ_PM} shows clearly that there is no systematic difference in the metallicity estimates between the two samples despite the fact they are derived with different methods/proxies. The two samples put together thus constitute a sample large enough to be considered as statistically significant.

Since it is known that metallicity is lower for higher SFR, and that the mean SFR is higher at higher redshift at a given stellar mass, \cite{2012PerezMontero} also studied if this can be the cause of the observed evolution of the MZR. The relation between the SFR-corrected metallicity, using equation 8 of \cite{2012PerezMontero}, and stellar mass for the combined MASSIV and zCOSMOS samples is very similar to the original mass-metallicity relation keeping a clear evolution with redshift. 



This means that this evolution is not due to any SFR-based selection effect as it is still there when correcting the metallicity by the SFR. To investigate further that the evolution is really due to the redshift and not to selection effects which can occur when we compare galaxy samples at different redshifts or with different selection criteria, it is interesting to look at the relation between the stellar mass and the nitrogen-to-oxygen abundance ratio, also called MNOR. If the MNOR shows the same evolution with redshift as the MZR it could mean that the evolution is real and that the FMR is not so fundamental, because the MNOR does not depend on SFR. Otherwise it could mean that the evolution we see is due to other effects such as metallicity computation issues. Computing the nitrogen-to-oxygen abundance ratio for high-$z$ individual galaxies is not an easy task as it requires the detection of several emission lines for each galaxies. For the MASSIV sample we computed the nitrogen-to-oxygen abundance ratio from the nitrogen-to-sulphur line ratio as explained in \cite{2009PerezContini}: 
\begin{equation}
\log (N/O) = 1.26 N2S2 - 0.86
\end{equation}
where $N2S2$ is the ratio between the flux of \nii\  and \sii. Considering the low $SNR$ of the integrated spectra of MASSIV galaxies we were not able to measure the \sii\ doublet for the majority of galaxies. We thus used stacked spectra for two bins of stellar mass and measure the flux of  \nii\  and \sii\ as described in \cite{2012Contini}. We find a value of $\log (N/O) = -0.74 \pm 0.25$ for $9.2 < \log(M_* / M_{\odot}) < 10.2$ and $\log (N/O) = -0.88 \pm 0.30$ for $10.2 < \log(M_* / M_{\odot}) < 11.2$.
We did not find any significant evolution of the MNOR between $z=0$ \citep{2012PerezMontero} and the one we tentatively derived at $z \sim 1.3$ with MASSIV stacked spectra. But with two points only at high redshift it is hard to reach any firm conclusion. 

\section{Discussion and conclusions}
\label{sect:conclusion}

In this paper we looked at some relations between fundamental parameters of 48 star-forming galaxies at $1<z<2$ for which we have been able to compute the metallicity from SINFONI data. 

First we investigated the relation between stellar mass and metallicity in order to quantify its evolution with redshift. 
As expected and in agreement with previous studies \citep{2009Queyrel}, MASSIV galaxies are in good agreement with the mass-metallicity relation defined by \cite{2012Yabe} for galaxies at the same median redshift ($z\sim 1.3$) and the same ranges of stellar mass and SFR. This relation seems however to depend on the range of SFR we look at. We found also that the median value of metallicity and the scattering around this relation is slightly different for isolated and interacting galaxies. Interacting galaxies have on average a lower metallicity than isolated ones, as observed also in the local universe \citep{2008MichelDansac}, and a lower dispersion around the MZR. It points out the importance of interactions in the chemical evolution of galaxies but it can also be partially explained by the difference of mass range for the two samples.

We then investigated the relations involving stellar mass, metallicity and SFR as done by \cite{2010Mannucci} and \cite{2010LaraLopez}. Indeed, these authors studied relations between these three parameters and found a fundamental plane which can be explained by an equilibrium model such as the one developed by \cite{2008Finlator} in which metallicity and SFR are strongly linked and depend mainly on gas inflows and outflows. Studies as the ones performed by \cite{2010LaraLopez} and \cite{2010Mannucci} are observational studies giving empirical relations and the origin and even the existence of such relations are still debated. 

On one hand, models predict that if galaxies are in a stable situation, described by simple instantaneous recycling models, the stellar mass and metallicity should not be depending on the SFR. Some recent studies such as \cite{2013Sanchez}, based on the CALIFA survey, find that metallicity does not depend on the SFR for a given mass range. They claim that the relations found by \cite{2010LaraLopez} and \cite{2010Mannucci} may be biased by the way SFR are computed and the apertures used in the SDSS and therefore by a selection effect depending on the redshift range studied. Nevertheless, in their recent paper, \cite{2013LaraLopez} selected their sample taking this possible bias into account and they imposed a low redshift limit depending on the fibre diameter of the instrument used for the observations. Despite this restricted selection, they found a fundamental plan in their data with slightly different parameters. 

On the other hand, a recent theoretical work has been performed \citep{2013Lilly} deriving an equation relating mass, metallicity and SFR from basic continuity equations and they found that SFR appears naturally as a second parameter in the mass-metallicity relation. Their equation gives a simple explanation of the fundamental metallicity relation and explains its independence with cosmic time. Other studies have been performed at higher redshift such as \cite{2013Belli} and seem to confirm the existence of the fundamental relation but also its invariance with redshift, but the samples used for such analysis are often small compare to samples in the local universe, as it is the case also for the MASSIV sample. 

To investigate further and check for any possible bias, we compared the MASSIV dataset to zCOSMOS galaxies at similar redshifts \citep{2012PerezMontero}. The MZR for the two samples are consistent and follows the expected relation for star-forming galaxies in the redshift range $z\sim 1-2$. In a second step we corrected the MZR by a function which depends on the SFR to remove the possible SFR-based selection effects, as performed in \cite{2012PerezMontero}. We still observe an evolution with redshift meaning that there should be also an evolution with redshift of the FMR. 

To conclude, the relations we found in our study could be consistent with the analysis of \cite{2010LaraLopez} and \cite{2010Mannucci} (if there is an evolution with redshift of their parameters) but given the size of our sample and the errors on the parameters it is difficult to confirm strictly the non-evolution of the FMR with redshift. 


\begin{acknowledgements}
We thank N. Bouch\'e and L. Michel-Dansac for useful discussions which help to improve this paper. We are grateful to the referee for his/her useful comments. This work has been partially supported by the CNRS-INSU and its Programme National Cosmologie-Galaxies (France) and by the French ANR grant ANR-07-JCJC-0009.
EPM acknowledges the project AYA2010-21887-C04-01 of Spanish Plan for Astronomy and Astrophysics.
\end{acknowledgements}

\newpage
\bibliographystyle{aa}
\bibliography{metal}

\end{document}